\documentclass[twocolumn,prd,nofootinbib,aps,prl,floats,floatfix,amsmath,amssymb,longbibliography,secnumarabic]{revtex4-1} %
\usepackage[english]{babel}

\usepackage[letterpaper,top=2cm,bottom=2cm,left=3cm,right=3cm,marginparwidth=1.75cm]{geometry}

\usepackage{amsmath}
\usepackage{graphicx}
\usepackage[colorlinks=true,citecolor=red]{hyperref}

\newcommand{\be}{\begin{equation}}
\newcommand{\ee}{\end{equation}}

\newcommand{\bea}{\begin{eqnarray}}
\newcommand{\eea}{\end{eqnarray}}
\newcommand{\nn}{\nonumber}

\begin{document}

\title{There is no 690 GeV resonance}
\author{James M.\ Cline}
\email{jcline@physics.mcgill.ca}
\affiliation{McGill University Department of Physics \& Trottier Space Institute, 3600 Rue University, Montr\'eal, QC, H3A 2T8, Canada}

\begin{abstract}
In a series of $\sim 30$ papers starting in 1991, it has been claimed that the Higgs field 
should be heavier than its now-measured value.  To reconcile this idea with reality, it was modified to the assertion that the Higgs field describes two physical degrees of freedom, one of which corresponds to a second Higgs particle with mass 690\,GeV.  Here I summarize the lack of theoretical and experimental evidence for these claims.

\end{abstract}

\maketitle

\section{Original comment}

Recently Ref.\ \cite{Consoli:2025lme} reiterated the claim, already made in Refs.\ \cite{Consoli:2025ezv,Rupp:2024llq,Consoli:2023zkq,Consoli:2023hnw,Consoli:2023hvz,Consoli:2022hks,Consoli:2022pfe,Consoli:2022lyl,Consoli:2021yjc,Consoli:2021hdo,Consoli:2020kip,Consoli:2020nwb,Cea:2019xdi}, that the Higgs field has an excited state with mass 690\,GeV.  This appears to be a modification of an earlier idea \cite{Castorina:2007ng,Cea:2003gp,Cea:2002zc,Cea:1999zu,Cea:1999kn,Cea:1998hy,Cea:1996af,Consoli:1995bz,Consoli:1995rc,Agodi:1994qv,Consoli:1993aw,Consoli:1992yg,Consoli:1991vj,Consoli:1991hy}, pursued by one of the same authors, that the Higgs mass could or should be above the 
perturbative unitarity limit $\sim 700$\,GeV,
as heavy as 2\,TeV, depending upon the year of publication.  The theoretical motivation for this prediction was the claim \cite{Branchina:1990ve,Castorina:1990br} that $\lambda\phi^4$ is not trivial, as is usually believed, but rather has a radiatively generated spontaneous symmetric phase (as predicted by the Coleman-Weinberg one-loop potential), in which it is asymptotically free.\footnote{The triviality of $\phi^4$ theory, long believed to be the case, was proven in Ref.\ \cite{Aizenman:2019yuo}.}

It was also claimed that the vacuum expectation value (VEV) of the scalar field gets renormalized by a different factor $Z_v$ than the fluctuations around the VEV, $Z_\phi$, so that the usual relation between the Higgs mass and the VEV is modified by a factor $\sqrt{Z_\phi/Z_v}$ which must be determined by lattice simulations, and predicts $m_h = 760\pm20\,$GeV \cite{Cea:2002zc}.

With the experimental discovery of the Higgs with mass $m_h = 125\,$GeV, one might have hoped for such claims to be put to rest, but a way to have one's cake and eat it too was found.
It somehow goes back to the aforementioned idea, that pure $\lambda\phi^4$ theory has spontaneous symmetry breaking {\it \`a la} Coleman-Weinberg, despite the usual reservations that the perturbative calculation leading to that result cannot be trusted.  The authors argue that now there are two mass scales in the potential: one is $m^2_h$, the curvature of the potential $V$ at its minimum,
and the other is $M_H^4 = \Delta V$, from the depth of the potential minimum, which was generated by radiative symmetry breaking.  It is not clear
why this extra scale should correspond to an additional propagating degree of freedom.

In order for a single field to describe two degrees of freedom, the propagator must have
two poles, which usually arises from a higher derivative action containing ghosts.  In the present case, the authors claim that nonperturbative effects generate the propagator
structure
\be
 G =  {i\over p^2 - M_H^2 A(p^2)} \label{eq1}
\ee
where $A$ is a function such that
$A(m_h^2) = m_h^2/M_H^2$ and $A(M_H^2) = 1$. The detailed form of $A(p)$ is not disclosed, so we are forced to guess.\footnote{Ref.\ \cite{Consoli:2025lme} says that this behavior was verified on the lattice in Ref.\ \cite{Consoli:2020nwb}, but that reference purports to show that the form of the inverse propagator is $(p^2-m_h^2)f(p)$, where $f(p)$ has the same properties as $A(p)$ in Eq.\ (\ref{eq1}).  This is puzzling since $f(p)$ corresponds to wave function renormalization, while $A(p)$ is the self-energy. }

It cannot be linear in $p^2$ since that would give $G = i/0$; hence the next simplest analytic possibility is quartic, $A = 1 + (p^4/M_H^4)(M_H^2/m_h^2 -1)$.  With this choice, we find for $m_h \ll M_H$
\be
    G \cong {-i M_H^2\over (p^2-M_H^2)(p^2-m_h^2)}\,,
\ee
which has the wrong sign for the heavy degree of freedom.  The heavy particle is a ghost, as expected from a theory with a higher-derivative Lagrangian.  The theoretical motivations for the ``resonance'' (unaptly named, since it is supposed to be coming from an elementary Higgs field, not a composite particle) are problematic.

\begin{figure*}[t]
\centerline{\includegraphics[width=0.9\columnwidth]{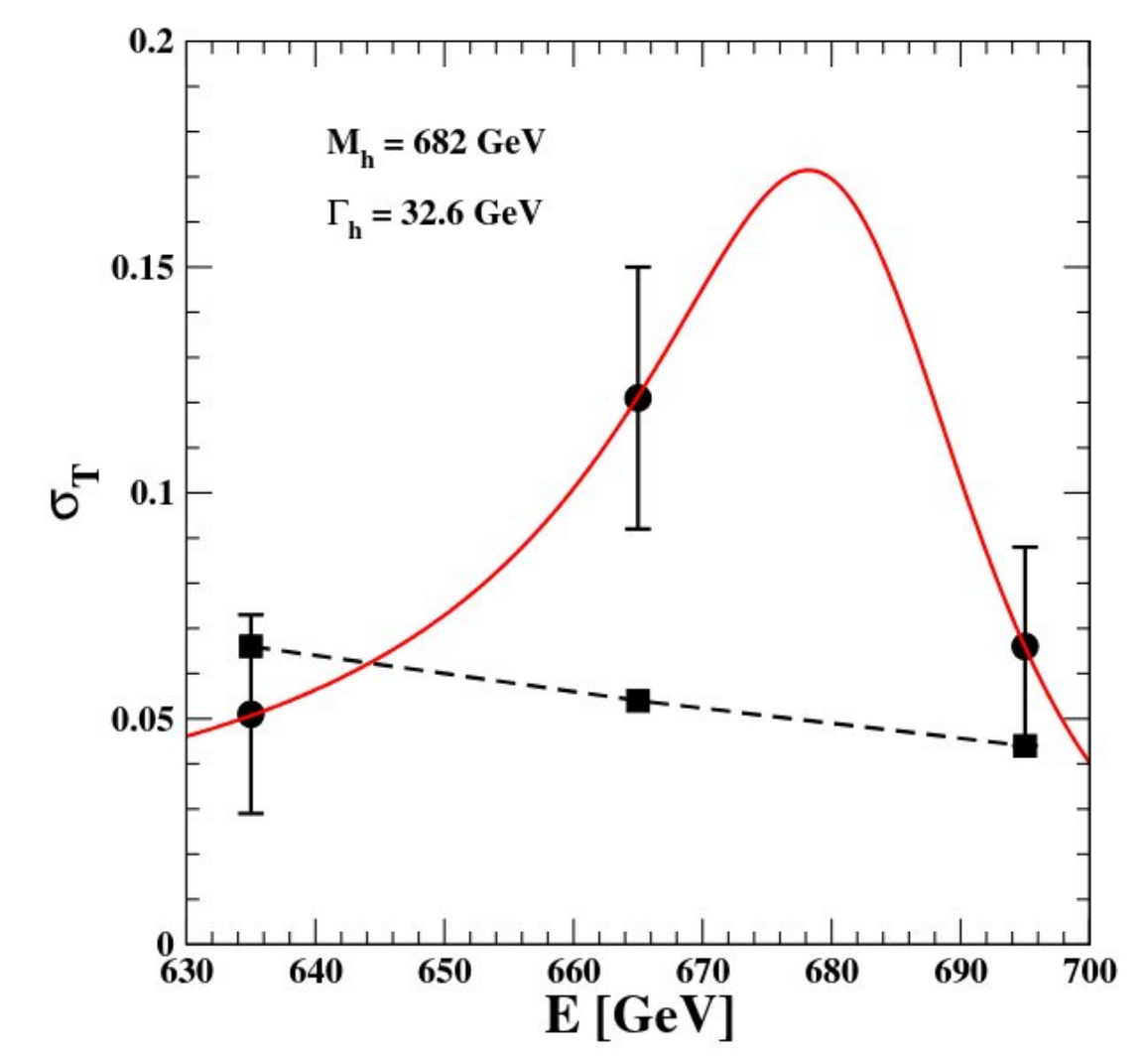}
\raisebox{-1em}{\includegraphics[width=1.2\columnwidth]{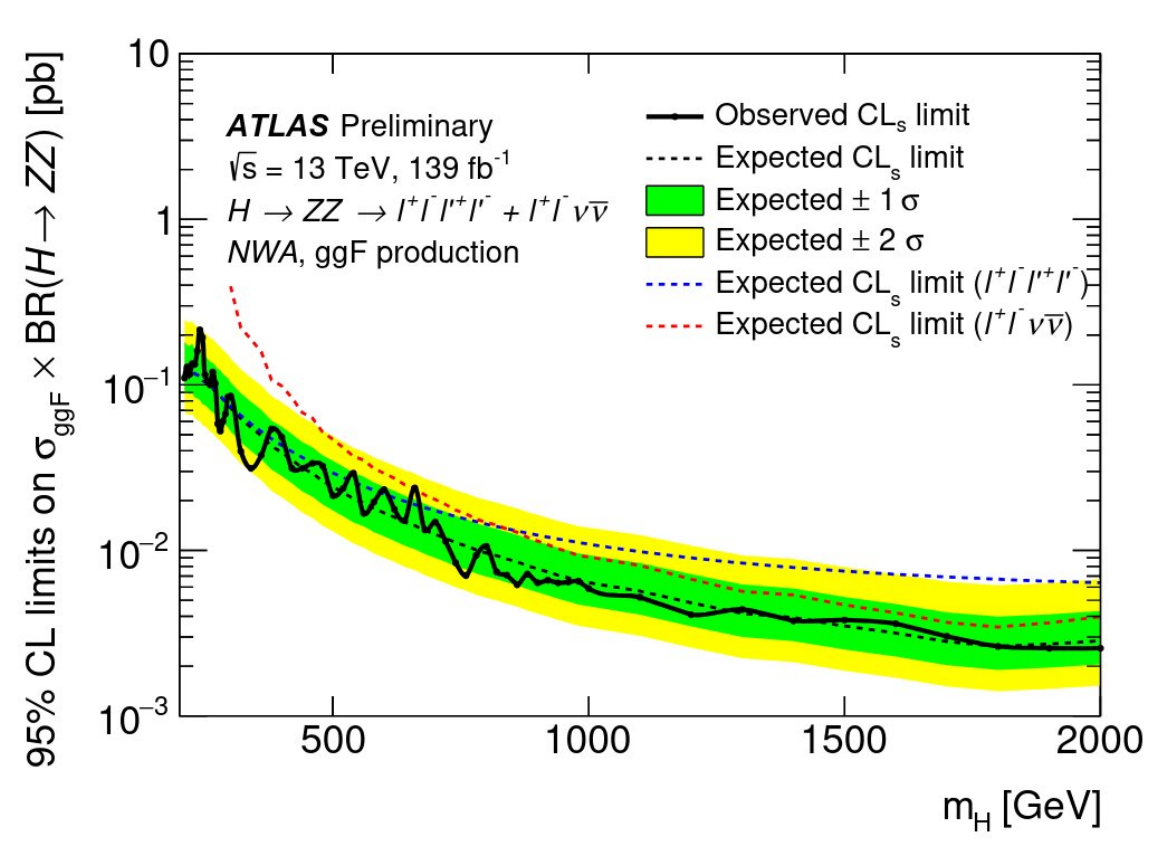}}}
\caption{Left: evidence for excess in $H\to ZZ\to 4\ell$ claimed by Ref.\ \cite{Consoli:2020kip}.  The red curve is their model prediction, points are derived from ATLAS results.  Right: ATLAS results for the same process \cite{ATLAS:2020tlo}, assuming ggF fusion.
} 
\label{fig:4lplot}
\end{figure*}

Let us turn then to the experimental evidence, which the LHC collaborations must have been very excited to discover.  In Ref.\ \cite{Consoli:2020kip} the authors discerned a bump in the ATLAS search \cite{ATLAS:2020tlo} for heavy resonances decaying to $ZZ \to 4\ell$
at $m_H\sim 700\,$GeV.  The authors note that $H$ should be dominantly produced through the gluon-gluon fusion (ggF) process, with negligible production from vector boson fusion (VBF).   Fig.\ \ref{fig:4lplot}
reproduces the main results from the two papers.
The ATLAS ggF limit has a 2-$\sigma$ excess at 662\,GeV, which receives no comment in the ATLAS paper, and only upper limits are quoted.

The CMS collaboration took note of Ref.\ \cite{Consoli:2020kip}'s prediction of an excess in this channel in their later search \cite{CMS:2024vps}. They also reported no significant excess.  Since the original suggestion \cite{Consoli:2020kip}, there have been an additional ten papers \cite{Consoli:2025lme,Consoli:2025ezv,Rupp:2024llq,Consoli:2023zkq,Consoli:2023hnw,Consoli:2023hvz,Consoli:2022hks,Consoli:2022lyl,Consoli:2022pfe,Consoli:2021yjc} by various combinations of the authors emphasizing the predicted excess, lest we should forget.  None of them are referred to by the experimental collaborations.  In fact, of the 44 citations to these papers, all but 11 are self-cites.  The authors find an equally convincing bump in
the $H\to hh$ channel, leading them to ``spell out a deﬁnite experimental signature of this resonance that is clearly visible in various LHC data.''  A Nobel prize is sure to follow.

\section{Response to authors' reply}

The authors of Ref.\ \cite{Consoli:2025lme} have kindly corrected my oversight about where to find an explicit formula for $A(p^2)$.  It is in Ref.\ \cite{Consoli:2023zkq}:\footnote{Following Ref.\ \cite{Okopinska:1995su} as cited by the authors, I find different coefficients for $J$ in the numerator and denominator, but this does not affect the following argument}
\be
    A(p^2) = {1-J\over 1 + J/2}\,,
    \label{Aeq}
\ee
where
\bea
    J &=& {\lambda\over 16\pi^2}\
    \Bigg(\ln{\Lambda^2\over M_H^2}\nn\\
&-&\int_0^1 dx\,\ln\left(1-x(1-x){p^2\over M_H^2}\right)\Bigg)\,,
\label{Jeq}
\eea
where $\Lambda$ is the ultraviolet cutoff, and 
I have translated their formula to the $(+,-,-,-)$
metric where the pole appears at $p^2 = M_H^2$ and not $-M_H^2$.\footnote{The authors seem to have overlooked this point since they write  the inverse propagator as $p^2 + M^2_H A(p^2)$ with $A=1$ in the $\lambda=0$ limit, while studying the behavior of $A(p^2)$ assuming $p^2>0$.}

The authors assume that $\lambda$ is renormalized
as $\lambda \sim 1/\ln\Lambda^2$ 
to compensate for the dependence on the cutoff as 
$\Lambda\to\infty$, but clearly one cannot take this limit while maintaining that the finite contribution has any relevance, so in fact they must keep the cutoff finite, such that the observed Higgs mass is supposed to be related to the postulated heavy Higgs mass by
\be
    m_h \approx {M_H\over \sqrt{\ln\Lambda^2/M_H^2}}\,.
\ee
Using $m_h=125$\,GeV and $M_H = 690\,$GeV, one obtains $\Lambda \sim 10^9\,$GeV, a scale which is seemingly never mentioned in their papers.  The logarithm in Eq.\ (\ref{Jeq}) can therefore be estimated as $\ln\Lambda^2/M_H^2 \approx 30.5$.

According to the authors, to exhibit the two poles of the propagator, we should evaluate
$A$ at $p^2=M_H^2$ and at $p^2\approx 0$.  The integral is simple to compute in these cases and gives
\be
    J \approx 30.5\,{\lambda\over 16\pi^2}\left( 1
    +\left\{\begin{array}{ll}-0.006, & p^2=M_H^2 \\
        \quad 0, & p^2=0\end{array}\right\}\right)\,.
\ee
In order to obtain $A(0) \approx (m_h/M_H)^2$, one must tune the value of $\lambda$ such that $J\approx 1-\epsilon$ when $p^2=0$, where $\epsilon = (3/2)m_h^2/M_H^2$.  But then one finds that $A(M_H^2) \approx (2/3)(\epsilon + 0.006)
\approx (2/3)\epsilon$, instead of $A(M_H^2) =1$ as claimed.  For such large values of the cutoff, 
the finite part of the loop integral has a negligible effect and the propagator has only one pole, as any graduate student learning quantum field theory would expect (and hopefully be taught)
in a theory of a single scalar field.

Eqs.\ (\ref{Aeq},\ref{Jeq}) are based on Ref.\ \cite{Okopinska:1995su}, which derived the corresponding result in the $O(N)$ version of $\lambda\Phi^4$ theory, using a nonperturbative variational method (Gaussian Effective Potential \cite{Stevenson:1985zy}, GEP), in which the present
authors take $N=1$ to apply it to the real Higgs field.  Ref.\ \cite{Okopinska:1995su} was written in response to another incorrect claim \cite{Brihaye:1985as}, namely that Goldstone's theorem is violated in the GEP analysis of the $O(2)$  $\lambda\Phi^4$ model.  Those authors found a massive Goldstone boson, which was proven to be massless in the correct treatment of the GEP in Ref.\ \cite{Okopinska:1995su}.  Hence, summarizing all of the claims, in this imaginary four-dimensional universe, Goldstone's theorem is violated, $\lambda\phi^4$ theory is asymptotically free,
and $\phi$ represents two physical degrees of freedom.

\bigskip
I thank Guy Moore for useful information, and the Niels Bohr International Academy for its generous hospitality.

\bibliographystyle{utphys}
\bibliography{sample}

\providecommand{\href}[2]{#2}\begingroup\raggedright\begin{thebibliography}{10}

\bibitem{Consoli:2025lme}
M.~Consoli, L.~Cosmai, F.~Fabbri, and G.~Rupp, ``{The 690 GeV scalar resonance},'' \href{http://arxiv.org/abs/2509.06479}{{\ttfamily arXiv:2509.06479 [hep-ph]}}.

\bibitem{Consoli:2025ezv}
M.~Consoli, L.~Cosmai, F.~Fabbri, and G.~Rupp, ``{Additional evidence of a new 690 GeV scalar resonance},'' \href{http://arxiv.org/abs/2501.03708}{{\ttfamily arXiv:2501.03708 [hep-ph]}}.

\bibitem{Rupp:2024llq}
G.~Rupp and M.~Consoli, ``{A New 700GeV Scalar in the LHC Data?},'' \href{http://dx.doi.org/10.31526/lhep.2024.515}{{\em LHEP} {\bfseries 2024} (2024) 515}, \href{http://arxiv.org/abs/2404.03711}{{\ttfamily arXiv:2404.03711 [hep-ph]}}.

\bibitem{Consoli:2023zkq}
M.~Consoli and G.~Rupp, ``{Second resonance of the Higgs field: motivations, experimental signals, unitarity constraints},'' \href{http://dx.doi.org/10.1140/epjc/s10052-024-13253-z}{{\em Eur. Phys. J. C} {\bfseries 84} no.~9, (2024) 951}, \href{http://arxiv.org/abs/2308.01429}{{\ttfamily arXiv:2308.01429 [hep-ph]}}.

\bibitem{Consoli:2023hnw}
M.~Consoli, L.~Cosmai, and F.~Fabbri, ``{Theoretical Arguments and Experimental Signals for a Second Resonance of the Higgs Field},'' \href{http://dx.doi.org/10.3390/universe9020099}{{\em Universe} {\bfseries 9} no.~2, (2023) 99}.

\bibitem{Consoli:2023hvz}
M.~Consoli, L.~Cosmai, and F.~Fabbri, ``{Hunting for the second Higgs resonance at LHC},'' \href{http://dx.doi.org/10.1393/ncc/i2024-24007-2}{{\em Nuovo Cim. C} {\bfseries 47} no.~1, (2023) 7}.

\bibitem{Consoli:2022hks}
M.~Consoli, L.~Cosmai, and F.~Fabbri, ``{Experimental signals for a new heavy resonance in the ATLAS and CMS data},'' \href{http://dx.doi.org/10.22323/1.414.0204}{{\em PoS} {\bfseries ICHEP2022} (2022) 204}.

\bibitem{Consoli:2022pfe}
M.~Consoli, ``{A second resonance of the Higgs field: Theoretical motivations and experimental signals},'' \href{http://dx.doi.org/10.1393/ncc/i2023-23019-8}{{\em Nuovo Cim. C} {\bfseries 46} no.~1, (2022) 19}.

\bibitem{Consoli:2022lyl}
M.~Consoli, L.~Cosmai, and F.~Fabbri, ``{Second resonance of the Higgs field: more signals from the LHC experiments},'' \href{http://arxiv.org/abs/2208.00920}{{\ttfamily arXiv:2208.00920 [hep-ph]}}.

\bibitem{Consoli:2021yjc}
M.~Consoli and L.~Cosmai, ``{Experimental signals for a second resonance of the Higgs field},'' \href{http://dx.doi.org/10.1142/S0217751X22500919}{{\em Int. J. Mod. Phys. A} {\bfseries 37} no.~14, (2022) 2250091}, \href{http://arxiv.org/abs/2111.08962}{{\ttfamily arXiv:2111.08962 [hep-ph]}}.

\bibitem{Consoli:2021hdo}
M.~Consoli, ``{A Hidden, Heavier Resonance of the Higgs Field},'' \href{http://dx.doi.org/10.5506/APhysPolB.52.763}{{\em Acta Phys. Polon. B} {\bfseries 52} no.~6-7, (2021) 763}, \href{http://arxiv.org/abs/2106.06543}{{\ttfamily arXiv:2106.06543 [hep-ph]}}.

\bibitem{Consoli:2020kip}
M.~Consoli and L.~Cosmai, ``{A resonance of the Higgs field at 700 GeV and a new phenomenology},'' \href{http://arxiv.org/abs/2007.10837}{{\ttfamily arXiv:2007.10837 [hep-ph]}}.

\bibitem{Consoli:2020nwb}
M.~Consoli and L.~Cosmai, ``{The mass scales of the Higgs field},'' \href{http://dx.doi.org/10.1142/S0217751X20501031}{{\em Int. J. Mod. Phys. A} {\bfseries 35} no.~20, (2020) 2050103}, \href{http://arxiv.org/abs/2006.15378}{{\ttfamily arXiv:2006.15378 [hep-ph]}}.

\bibitem{Cea:2019xdi}
P.~Cea, M.~Consoli, and L.~Cosmai, ``{Two mass scales for the Higgs field?},'' \href{http://arxiv.org/abs/1912.00849}{{\ttfamily arXiv:1912.00849 [hep-ph]}}.

\bibitem{Castorina:2007ng}
P.~Castorina, M.~Consoli, and D.~Zappala, ``{An Alternative heavy Higgs mass limit},'' \href{http://dx.doi.org/10.1088/0954-3899/35/7/075010}{{\em J. Phys. G} {\bfseries 35} (2008) 075010}, \href{http://arxiv.org/abs/0710.0458}{{\ttfamily arXiv:0710.0458 [hep-ph]}}.

\bibitem{Cea:2003gp}
P.~Cea, M.~Consoli, and L.~Cosmai, ``{Indications on the Higgs boson mass from lattice simulations},'' \href{http://dx.doi.org/10.1016/S0920-5632(03)02711-7}{{\em Nucl. Phys. B Proc. Suppl.} {\bfseries 129} (2004) 780--782}, \href{http://arxiv.org/abs/hep-lat/0309050}{{\ttfamily arXiv:hep-lat/0309050}}.

\bibitem{Cea:2002zc}
P.~Cea, M.~Consoli, and L.~Cosmai, ``{New indications on the Higgs boson mass from lattice simulations},'' \href{http://arxiv.org/abs/hep-ph/0211329}{{\ttfamily arXiv:hep-ph/0211329}}.

\bibitem{Cea:1999zu}
P.~Cea, M.~Consoli, and L.~Cosmai, ``{Large rescaling of the Higgs condensate: Theoretical motivations and lattice results},'' \href{http://dx.doi.org/10.1016/S0920-5632(00)91767-5}{{\em Nucl. Phys. B Proc. Suppl.} {\bfseries 83} (2000) 658--660}, \href{http://arxiv.org/abs/hep-lat/9909055}{{\ttfamily arXiv:hep-lat/9909055}}.

\bibitem{Cea:1999kn}
P.~Cea, M.~Consoli, L.~Cosmai, and P.~M. Stevenson, ``{Further lattice evidence for a large rescaling of the Higgs condensate},'' \href{http://dx.doi.org/10.1142/S0217732399001760}{{\em Mod. Phys. Lett. A} {\bfseries 14} (1999) 1673--1688}, \href{http://arxiv.org/abs/hep-lat/9902020}{{\ttfamily arXiv:hep-lat/9902020}}.

\bibitem{Cea:1998hy}
P.~Cea, M.~Consoli, and L.~Cosmai, ``{First lattice evidence for a nontrivial renormalization of the Higgs condensate},'' \href{http://dx.doi.org/10.1142/S0217732398002515}{{\em Mod. Phys. Lett. A} {\bfseries 13} (1998) 2361--2368}, \href{http://arxiv.org/abs/hep-lat/9805005}{{\ttfamily arXiv:hep-lat/9805005}}.

\bibitem{Cea:1996af}
P.~Cea, L.~Cosmai, M.~Consoli, and R.~Fiore, ``{Lattice effective potential of ($\lambda phi^{4)}$ in four-dimensions: Nature of the phase transition and bounds on the Higgs mass},'' \href{http://arxiv.org/abs/hep-th/9603019}{{\ttfamily arXiv:hep-th/9603019}}.

\bibitem{Consoli:1995bz}
M.~Consoli and Z.~Hioki, ``{Indications on the Higgs boson mass from the LEP data},'' \href{http://dx.doi.org/10.1142/S0217732395002404}{{\em Mod. Phys. Lett. A} {\bfseries 10} (1995) 2245--2252}, \href{http://arxiv.org/abs/hep-ph/9505249}{{\ttfamily arXiv:hep-ph/9505249}}.

\bibitem{Consoli:1995rc}
M.~Consoli and Z.~Hioki, ``{Remarks on the value of the Higgs mass from the present LEP data},'' \href{http://dx.doi.org/10.1142/S0217732395000910}{{\em Mod. Phys. Lett. A} {\bfseries 10} (1995) 845--852}, \href{http://arxiv.org/abs/hep-ph/9503288}{{\ttfamily arXiv:hep-ph/9503288}}.

\bibitem{Agodi:1994qv}
A.~Agodi, G.~Andronico, and M.~Consoli, ``{Lattice $\phi^4$ in four-dimensions effective potential giving spontaneous symmetry breaking and the role of the Higgs mass},'' \href{http://dx.doi.org/10.1007/BF01556370}{{\em Z. Phys. C} {\bfseries 66} (1995) 439--452}, \href{http://arxiv.org/abs/hep-lat/9410001}{{\ttfamily arXiv:hep-lat/9410001}}.

\bibitem{Consoli:1993aw}
M.~Consoli and P.~M. Stevenson, ``{Resolution of the $\lambda \phi^4$ puzzle and a 2-TeV Higgs boson},'' \href{http://arxiv.org/abs/hep-ph/9303256}{{\ttfamily arXiv:hep-ph/9303256}}.

\bibitem{Consoli:1992yg}
M.~Consoli, ``{Spontaneous symmetry breaking and the Higgs mass},'' \href{http://dx.doi.org/10.1016/0370-2693(93)91108-Y}{{\em Phys. Lett. B} {\bfseries 305} (1993) 78--83}.

\bibitem{Consoli:1991vj}
M.~Consoli, ``{Large Higgs mass, triviality and asymptotic freedom},'' in {\em {Conference on Gauge Theories - Past and Future, in Honor of the 60th Birthday of M.J.G. Veltman}}, pp.~81--91.
\newblock 1991.

\bibitem{Consoli:1991hy}
M.~Consoli, V.~Branchina, P.~Castorina, and D.~Zappala, ``{Experimental constraints about the top and theoretical ideas about the Higgs},'' in {\em {26th Rencontres de Moriond: Electroweak Interactions and Unified Theories}}, pp.~131--139.
\newblock 1991.

\bibitem{Branchina:1990ve}
V.~Branchina, P.~Castorina, M.~Consoli, and D.~Zappala, ``{Nontriviality of spontaneously broken $\lambda \phi^4$ theories},'' \href{http://dx.doi.org/10.1103/PhysRevD.42.3587}{{\em Phys. Rev. D} {\bfseries 42} (1990) 3587--3590}.

\bibitem{Castorina:1990br}
P.~Castorina and M.~Consoli, ``{Asymptotic Freedom of Massless $\lambda \phi^4$ Theories},'' \href{http://dx.doi.org/10.1016/0370-2693(90)91968-H}{{\em Phys. Lett. B} {\bfseries 235} (1990) 302--304}.

\bibitem{Aizenman:2019yuo}
M.~Aizenman and H.~Duminil-Copin, ``{Marginal triviality of the scaling limits of critical 4D Ising and $\phi_4^4$ models},'' \href{http://dx.doi.org/10.4007/annals.2021.194.1.3}{{\em Annals Math.} {\bfseries 194} no.~1, (2021) 163}, \href{http://arxiv.org/abs/1912.07973}{{\ttfamily arXiv:1912.07973 [math-ph]}}.

\bibitem{ATLAS:2020tlo}
{\bfseries ATLAS} Collaboration, G.~Aad {\em et~al.}, ``{Search for heavy resonances decaying into a pair of Z bosons in the $\ell ^+\ell ^-\ell '^+\ell '^-$ and $\ell ^+\ell ^-\nu {{\bar{\nu }}}$ final states using 139 $\mathrm {fb}^{-1}$ of proton{\textendash}proton collisions at $\sqrt{s} = 13\,$TeV with the ATLAS detector},'' \href{http://dx.doi.org/10.1140/epjc/s10052-021-09013-y}{{\em Eur. Phys. J. C} {\bfseries 81} no.~4, (2021) 332}, \href{http://arxiv.org/abs/2009.14791}{{\ttfamily arXiv:2009.14791 [hep-ex]}}.

\bibitem{CMS:2024vps}
{\bfseries CMS} Collaboration, ``{Search for heavy scalar resonances decaying to a pair of Z bosons in the 4-lepton final state at 13 TeV}.'' \url{https://inspirehep.net/files/8f0aee7eb2c891e450654b419f9a9148}, 2024.
\newblock CMS-PAS-HIG-24-002.

\bibitem{Okopinska:1995su}
A.~Okopinska, ``{Goldstone bosons in the Gaussian approximation},'' \href{http://dx.doi.org/10.1016/0370-2693(96)00198-0}{{\em Phys. Lett. B} {\bfseries 375} (1996) 213--216}, \href{http://arxiv.org/abs/hep-th/9508087}{{\ttfamily arXiv:hep-th/9508087}}.

\bibitem{Stevenson:1985zy}
P.~M. Stevenson, ``{The Gaussian Effective Potential. 2. Lambda phi**4 Field Theory},'' \href{http://dx.doi.org/10.1103/PhysRevD.32.1389}{{\em Phys. Rev. D} {\bfseries 32} (1985) 1389--1408}.

\bibitem{Brihaye:1985as}
Y.~Brihaye and M.~Consoli, ``{SPONTANEOUS SYMMETRY BREAKING IN AN O(2) INVARIANT SCALAR THEORY},'' \href{http://dx.doi.org/10.1016/0370-2693(85)91209-2}{{\em Phys. Lett. B} {\bfseries 157} (1985) 48--52}.

\end{thebibliography}\endgroup

\end{document}